\documentclass[12pt]{article}

\usepackage{scicite}
\usepackage[utf8]{inputenc}
\usepackage[T1]{fontenc}
\usepackage[english]{babel}
\usepackage{parskip}
\usepackage{graphicx}
\usepackage{epstopdf}
\usepackage{breakcites}
\usepackage{array}
\usepackage{lscape}

\newenvironment{sciabstract}{%
\begin{quote} \bf}
{\end{quote}}

\newcounter{lastnote}

\title{Simulating feedback mechanisms in patient flow and return visits in an emergency department} 

\author
{Christian Michel Sørup,$^{1\ast}$ Daniel Alberto Sepúlveda Estay,$^{2}$ Peter Jacobsen,$^{2}$ \\ 
Philip Dean Anderson $^{3}$\\
\\
\normalsize{$^{1}$Rigshospitalet, Department of Improvement}\\
\normalsize{Blegdamsvej 9, 2200 Copenhagen (Denmark)}\\
\normalsize{$^{2}$Technical University of Denmark, Management Science} \\
\normalsize{Akademivej, bldg 358, 2800 Kongens Lyngby (Denmark)} \\
\normalsize{$^{1}$Brigham and Women's Hospital, Department of Emergency Medicine}\\
\normalsize{75 Francis St, Boston, MA 02115 (USA)}\\
\\
\normalsize{$^\ast$To whom correspondence should be addressed; E-mail:  christian.michel.soerup@regionh.dk}
}
\date{}

\begin{document} 
\baselineskip16pt
\maketitle 

\begin{sciabstract}
Emergency department (ED) crowding has been an increasing problem worldwide. Prior
research has identified factors that contribute to ED crowding. However, the
relationships between these remain incompletely understood. This study's objective was
to analyse the effects of initiating a local protocol to alleviate crowding situations
at the expense of increasing returning patients through the development of a system
dynamics (SD) simulation model. The SD study is from an academic care hospital in Boston, MA. 
Data sources include direct observations, semi-structured interviews, archival data from 
October 2013, and peer-reviewed literature from the domains of emergency medicine and 
management science. The SD model shows interrelations between inpatient capacity restraints 
and return visits due to potential premature discharges. The model reflects the vulnerability 
of the ED system when exposed to unpredicted increases in demand. Default trigger values for 
the protocol are tested to determine a balance between increased patient flows and the number 
of returning patients. Baseline simulation runs for generic variables assessment showed high 
leverage potential in bed assignment- and transfer times.

A thorough understanding of the complex non-linear behaviour of causes and effects of ED crowding 
is enabled through the use of SD. The vulnerability of the system lies in the crucial interaction 
between the physical constraints and the expedited patient flows through protocol activation. 
This study is an example of how hospital managers can benefit from virtual scenario testing 
within a safe simulation environment to immediately visualise the impacts of policy adjustments. 
\end{sciabstract}

\section{Introduction}
\label{intro}
Emergency department (ED) crowding has for decades been acknowledged as an international problem
compromising health outcomes \cite{Pines2011}. Hospitals have been cutting down on the number of inpatient beds primarily for budget reasons while the demand for emergency care services is on the rise \cite{Cameron2006}. The
American College of Emergency Physicians (ACEP) updated their position statement in February 2013
along with various policy recommendations aimed at mitigating crowding situations \cite{AmericanCollegeofEmergencyPhysicians2013}. 
Despite the recommendations, crowded EDs were still a growing problem. In 2006, the Institute of Medicine published their report entitled \textit{Hospital-Based
Emergency Care at the Breaking Point} with a warning that the current system is not sustainable \cite{InstituteofMedicine2006}.
In response, The Joint Commission stated that every hospital needs a plan for dealing with crowding 
in their ED \cite{Fee2007}. So far, most US hospitals do not seem to be fully capable to manage large-scale influx of patients, for instance as a result of terror acts. The same hospitals however play a critical role in both identifying and responding to terrorist attacks \cite{Matusitz2007}. Today, congested EDs remain an increasing problem in which causative factors remain to be fully elucidated \cite{Derlet2000}. While this paper refer to crowding in the ED, the authors do recognise
that crowding is a symptom of a system-wide problem associated with likely deterioration of health outcomes for patients across departments
\cite{Viccellio2013}. 

Simulation studies are increasingly acknowledged as a means to better understand the complex interconnections causing ED
crowding \cite{Paul2010}. Many of these are discrete event studies while system dynamics studies are rarer. Examples of the latter include discharging
strategies \cite{Wong2010} and the competition for inpatient beds between ED- and elective patients \cite{Lane2000}. Because of the problem's 
dynamic complexity, time delays through non-linear behaviour and feedback mechanisms, ED crowding seems apt for a system dynamics analysis. The problem is a
classic example of ''policy resistance'' when attempted interventions show poor results or even contradict intended effects. According to Sterman,
the problem persists due to an imperfect perception of what constitutes the problem \cite{Sterman2000}. This study's focus is twofold. First, a local protocol
('Code Help') for responding to ED crowding at a major tertiary level one
teaching hospital in Boston, MA is explored for its system-wide effect on patient flow. When predetermined criteria for maximum levels of ED boarders- and census are met, the protocol can be activated so that inpatient wards are bound to receive ED boarders by either reallocating current inpatients 
or discharge faster than normally. An ED boarder is defined as a patient who remains in the ED after having been admitted to an inpatient ward but
has not yet been transferred there \cite{AmericanCollegeofEmergencyPhysicians2013}. ED census denotes the total number of patients present in the ED at any given time. Congestion can dissolve but comes at the expense of a higher risk of premature
discharges \cite{Rising2013, Crawford2014, Jack2009}. Second, the model considers 1) inpatient capacity \cite{Trzeciak2003},
2) bed assignment time \cite{Rechel2010}, 3) transfer time \cite{Beniuk2012}, and 4) number of incoming elective patients
\cite{Morris2012} as potential levers for ED crowding. 

\section{Method and Materials}
\label{method}

\subsection{Setting}
\label{setting}
The SD model shows the admission process at the case hospital. Most patients go through the ED
in order to get a preliminary diagnosis and are either fully treated in the ED or, if further treatment is
required, admitted to an inpatient ward. 35 \% of the ED patients are admitted. This paper uses the notion of ED patients meaning essentially all acute patients. 
This is true for most hospitals today but it is recognised that some hospitals may still receive acute patients directly 
at inpatient wards.  

\subsection{Model construction}
\label{model construct}
Three main elements constitute the model. These are 1) the ED admission process, 2) two feedback mechanisms
controlling the available bed capacity and the effects of initiating the protocol accordingly, and 3) the competing
requirement for inpatient beds by elective patients.
For the admission process, all patients that enter the ED undergo an initial ESI triage assessment
\cite{Shelton2010}. Patients then receive medical evaluation and a preliminary diagnosis is established.

In the SD model, treatment is either finished in the ED or specialised treatment is needed elsewhere in the hospital. 
Patients requiring admission need to have an inpatient bed assigned. At the case study hospital, this happens via an
electronic request initiated from the ED where the receiving ward can choose either to accept the patient or request further
discussion with the ED team. This solution was implemented as part of a Lean initiative to avoid cumbersome telephonic
consultation. Next, patient admission is restricted by the bed occupancy rate to avoid assigning more beds than available. 
In case of ED crowding, the protocol can be activated if the following three criteria are met:

\begin{enumerate}
\item There are 10 or more ED boarders present,
\item ED census is 54 patients or more, and
\item Criteria 1 and 2 are met for two consecutive hours. 
\end{enumerate}

\begin{figure*}[ht!]
  \includegraphics[width=1\textwidth]{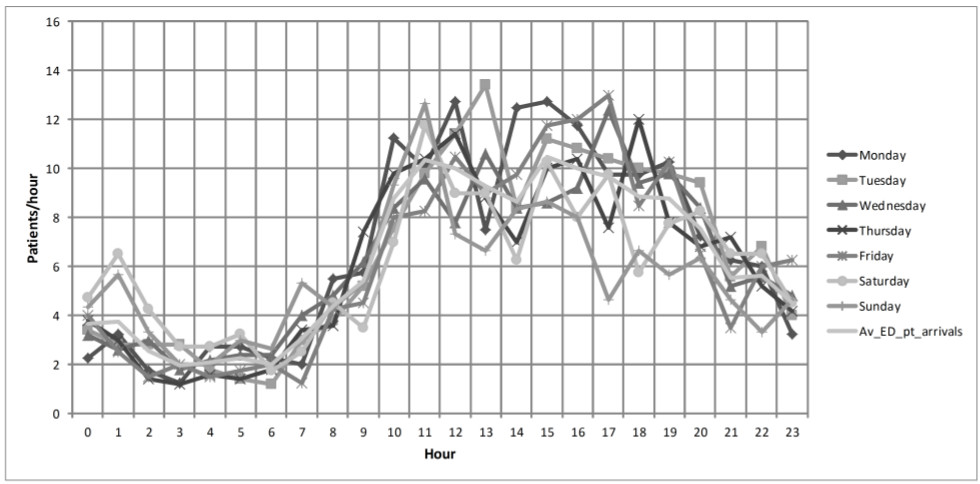}
\caption{Hourly ED patient arrival pattern following Lane et al., 2000}
\label{fig:1}       % Give a unique label
\end{figure*}

Upon activation, patients are either relocated or discharged faster from the inpatient wards, thereby increasing capacity for
incoming patients. Hence, the patient flow is increased and congestion eased. The expense is an increased risk
of discharging patients prematurely which could result in more patients returning unexpectedly \cite{Rising2013,Crawford2014,Jack2009}.\smallskip
The model SD was built using daily historical data from October 2013 stemming from the case study site. Inputs to the model
include day of the week averages for patient arrivals as modelled by Lane et al., 2000 (shown in Figure \ref{fig:1}), ED census, ED patients that await inpatient bed assignment, and ED boarders. The registration frequency was every ten minutes. With regards to how the ED patient arrival input signal is modelled, averaging the weekdays does imply losing some common variation. With a more advanced analytical input function, this variance could be included in future studies but is deemed beyond scope by the authors. Vensim DSS version 6.3 
was utilised to construct the model. Complete model documentation can be found in the appendix. During the testing of the model, it was seen that the perturbations upon a stable model 
dissipate within 72 hours. With this in mind, one week was adopted as the time frame for this analysis.

Upon activation, patients are either relocated or discharged faster from the inpatient wards, thereby increasing capacity for
incoming patients. Hence, the patient flow is increased and congestion eased. The expense is an increased risk
of discharging patients prematurely which could result in more patients returning unexpectedly \cite{Rising2013,Crawford2014,Jack2009}.\smallskip
The model SD was built using daily historical data from October 2013 stemming from the case study site. Inputs to the model
include day of the week averages for patient arrivals as modelled by Lane et al., 2000 (shown in Figure \ref{fig:1}), ED census, ED patients that await inpatient bed assignment, and ED boarders. The registration frequency was every ten minutes. With regards to how the ED patient arrival input signal is modelled, averaging the weekdays does imply losing some common variation. With a more advanced analytical input function, this variance could be included in future studies but is deemed beyond scope by the authors. Vensim DSS version 6.3 
was utilised to construct the model. Complete model documentation can be found in the appendix. During the testing of the model, it was seen that the perturbations upon a stable model 
dissipate within 72 hours. With this in mind, one week was adopted as the time frame for this analysis.

\subsection{Model structure}
\label{model structure}
\begin{figure*}
  \includegraphics[width=1\textwidth]{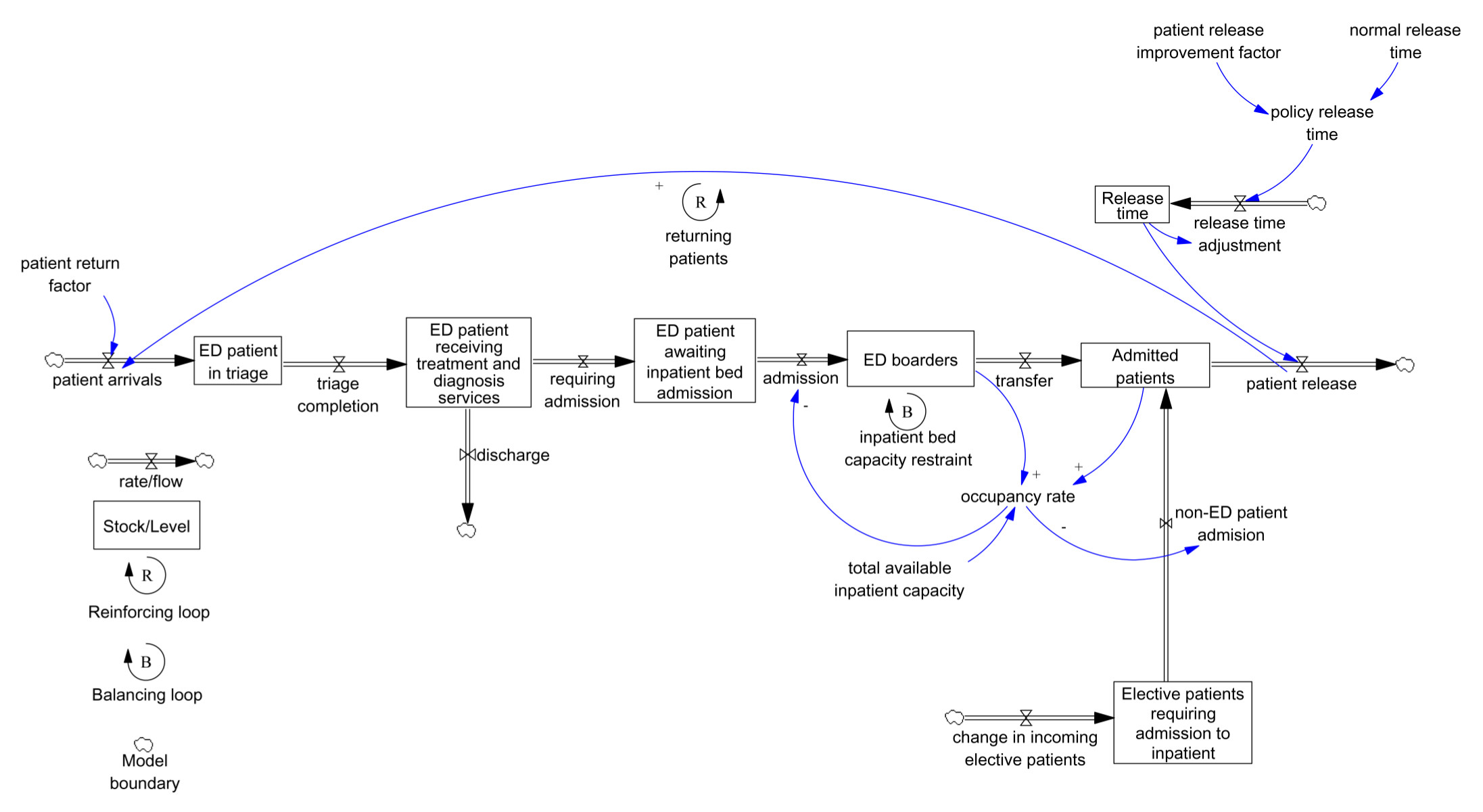}
\caption{Reduced presentation of the system dynamics model with auxiliary variables omitted. Patients arrive to the ED based on an
average repeated diurnal pattern and stay in the ED until either discharged or admitted. Patients requiring admission first need an inpatient
bed assigned, becoming ED boarders when the inpatient bed is assigned and are then transferred out of the ED. The feedback mechanisms, denoted R
and B respectively, and the inflow of elective patients are also presented. }
\label{fig:2}       % Give a unique label
\end{figure*}

A simplified presentation of the main SD model is shown in Figure \ref{fig:2}. The main flow presents the ED patient journey with
special emphasis on the admitted patients. The model includes two sub views (not shown). The first sub view is an exogenous mechanism that can
generate an inflow of patients in a user specified pulse, step, ramp, or sine wave function. This inflow is directly connected to the patient
arrivals inflow. The second sub view introduces a memory function representing the third criteria of the protocol, being thresholds for both ED
census- and boarders respectively are met for two consecutive hours.

The balancing feedback loop restricts the amount of patients to be admitted. The auxiliary variable 'occupancy rate'
reflects the availability of inpatient beds. With a higher occupancy rate it will become
increasingly difficult to admit new ED patients.

The reinforcing feedback loop explicitly models the return flow of admitted patients that have been discharged. Here,
a fraction of discharged patients need repeated treatment upon discharge within 72 hours. This fraction increases
when the protocol is activated because the discharging of patients at a rate faster than normal will result in a higher 
probability of a patient developing the need of readmission. The upper right corner in Figure 2 models the release time, 
equivalent to a patient's total time as inpatient. Activation of the protocol impacts the policy release time 
and the protocol's effect will show after a 30 minutes delay. The before mentioned fraction of patients returning to the 
ED is based on estimates found in literature \cite{Rising2013,Crawford2014}. 

The structure of the model finally considers four independent input parameters, namely total available inpatient beds, Bed Assignment Time,
Transfer Time, and Mean Elective Patients. These are considered as exogenous variables to the model. Also, the SD model considers four dependent output variables upon which system performance is evaluated. These are 1) occupancy, 2) ED Census, 3) ED Boarders, and 4) number of admitted elective patients.

\subsection{Testing procedure}
\label{sens_test_procedure}

After model validation, several simulations were run with various parameter settings. Adjusting the trigger threshold values to see the
effects on ED crowding level and returning patients was of primary interest. In order for the protocol to become activated
temporarily, the baseline model was stressed by introducing a sudden surge of patients arriving within three hours. In the baseline scenario, the model is 
simulated according to normal system conditions and no range variation is considered in the input parameters. There is no
activation of the Code Help protocol in the baseline model. What distinguishes the stressed model from the baseline model is a temporary activation of the Code Help 
protocol due to an exogenous stimulus. This is provoked by introducing a temporary surge of incoming ED patients with a subsequent elevation in the fraction of admitted ED patients.   
The admission fraction had to be increased in the period of the surge after a small delay corresponding to patients having undergone initial treatment. Trigger values were
changed separately to determine the subsequent effects on the outcome variables. First, the ED census trigger was allowed to vary from 40
to 68 (default = 54). Next, the ED boarder trigger would vary from 7 to 13 (default = 10). The ED census trigger criterion is met on a
regular basis even in the baseline simulation. Based on the semi-structured interviews conducted with staff physicians, the main indicator
of ED crowding is the accumulation of ED boarders. It was expected that altering this trigger would have the largest effect on the
system. The simulation model developed offers ED decision makers insights into other potential policy changes. Other parameters analysed include
inpatient capacity, bed assignment time, transfer time, and number of incoming elective patients.

\section{Results}

\subsection{Model validation}
\label{model_validate_results}

If the model is deemed sufficiently valid then the stakeholders are more likely to be assured that the system is replicated to
satisfaction. Tenured physicians were iteratively consulted to qualitatively agree on the correctness of the process, feedback
mechanisms, and assumptions modelled. Quantitative testing of the model was done through comparison of model output to historical
data of ED census from October 2013. As advocated by Sterman, replication of historical data does not provide sufficient evidence for a
robust model \cite{Sterman2002}. Therefore, the developed SD model was subject to a number of tests to ensure feasible behaviour
under extreme conditions. 

\begin{figure}[htb]
  \includegraphics[width=1\textwidth]{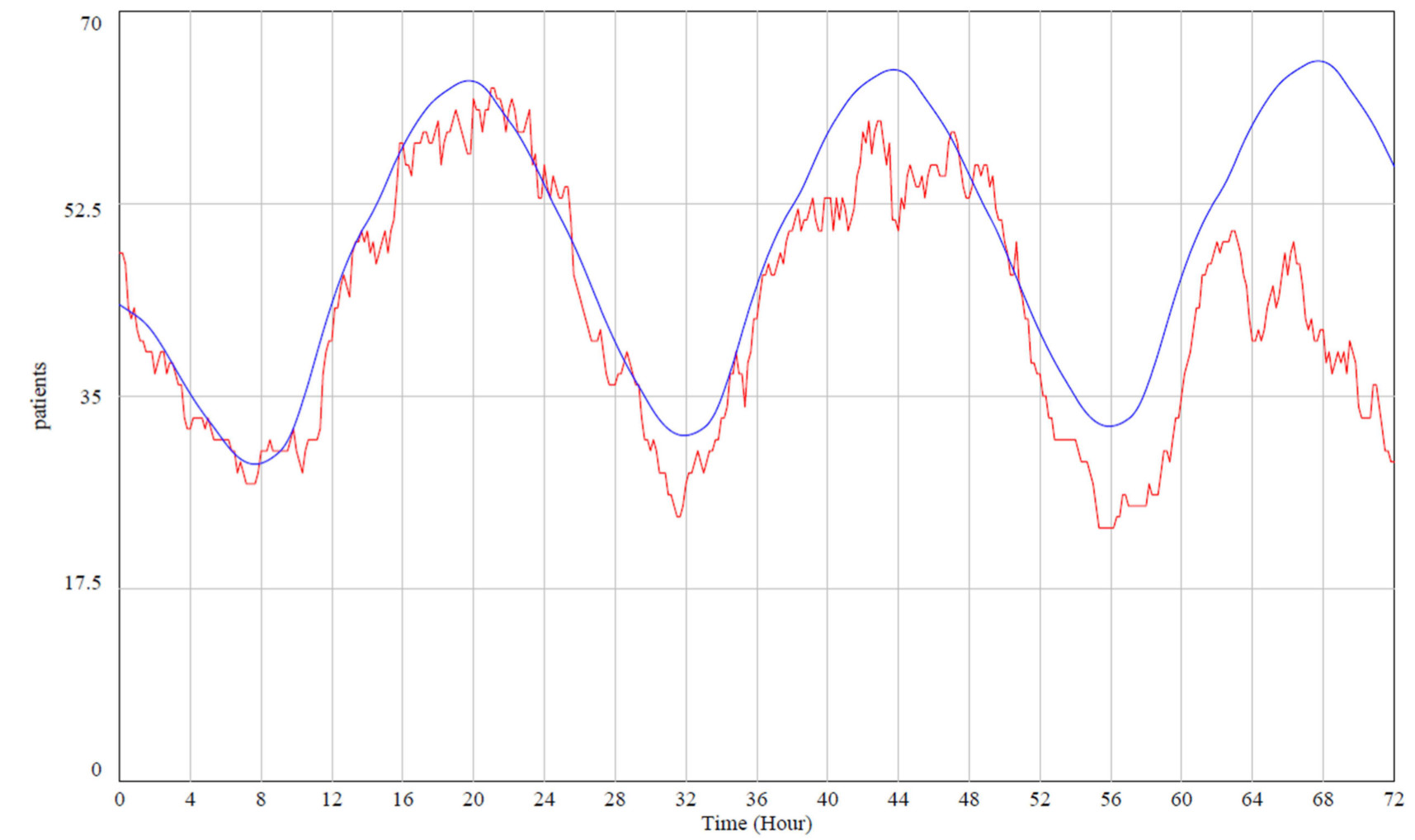}
\caption{Comparison of ED census data and simulation output for first 72 hours (October 1$^{st}$ to 3$^{rd}$, 2013). The smooth graph shows the
simulation output while the jagged graph shows imported historical data.}
\label{fig:3}       % Give a unique label
\end{figure}

Quantitative validation was done by comparing the simulation output for ED census to the acquired historical data reflecting the actual ED
census throughout October 2013. Generally, there is a good fit between the two graphs (see Figure \ref{fig:3}). Extreme tests were
performed as to determine system behaviour when stressed. Confidence in the model's robustness was established between authors and study
site physicians involved in the development of the model. 

\subsection{Sensitivity testing}
\label{sens_test_results}
As the model can replicate actual behaviour adequately, the study proceeds to testing how the ED census- and ED
boarders trigger values can change the model's behaviour. Such changes correspond to an earlier or later activation 
of the protocol. Effects of parameter changes are measured on the occupancy rate, admitted patients, ED census, 
and ED boarders. Testing the settings for each trigger is conducted in isolation keeping all other parameters constant.
The baseline model investigates other relevant generic parameters that are potential levers for alleviating ED
congestion. 

The sensitivity analysis was performed through 200 Monte Carlo simulations of the model for each of the input variables, considering uniform distributions for these inputs. The result of these random input parameters was then recorded for each of the selected output variables. The results of this simulation process have been summarised in the data contained in Tables \ref{tab:2} and \ref{tab:3} \cite{Rahmandad2012}.

\begin{landscape}
\begin{table*}

\caption{Stressed model sensitivity test results for protocol trigger values. The percentages of deviation to the default value are in parentheses. The values have been extracted at the time where the three graphs (base, maximum, and minimum) show the biggest variance. Most leverage potential is found in the largest percentages deviation. Note no effect on
varied ED census trigger due to the threshold being frequently exceeded}
\label{tab:1}       % Give a unique label

\begin{tabular}{l|r|r|r||r r r|}
%\begin{tabular}{l|p{1.3cm}|p{1.3cm}|p{1.3cm}||p{1cm} p{1cm} p{1cm}|}
\cline{2-7}
\multicolumn{1}{l|}{} & \multicolumn{3}{c||}{Trigger 1: ED boarders} & \multicolumn{3}{c|}{Trigger 2: ED census} \\
\cline{2-7} 
 & 7 \newline (Min.) & 10 \newline(Base) & 13 \newline (Max.)& 48 \newline (Min.) & 54 \newline (Base) & 60\newline (Max.)\\ \hline 
Occupancy & 0.8971 \newline(-2.54\%) & 0.9205 & 0.926 (+0.59\%) & \multicolumn{3}{c|}{no independent effect} \\
\textit{time, max diff} & & 51 & & & & \\ \hline
ED census & 64.34 (+5.47\%) & 61.02 & 60.06 (-1.57\%) & \multicolumn{3}{c|}{no independent effect} \\
\textit{time, max diff} & & 67.5 & & & & \\ \hline
ED boarders & 6.503 (+5.69\%) & 6.153 & 6.052 (-1.64\%) & \multicolumn{3}{c|}{no independent effect} \\
\textit{time, max diff} & & 70.125 & & & & \\ \hline
Admit elect patients & 442 (-2.58\%) & 453.7 & 456.6 (+0.64\%) & \multicolumn{3}{c|}{no independent effect} \\
\textit{time, max diff} & & 51 & & & & \\ \hline

\end{tabular}
\end{table*}

%table divided by Daniel
\begin{table*}
\caption{Baseline sensitivity test results for inpatient beds and bed assignment time}
\label{tab:2}       % Give a unique label
\begin{tabular}{l | r | r | r || r| r | r |}
%\begin{tabular}{r | p{1.4cm} | p{1cm} | p{1.4cm} || p{1.5cm}| p{1.5cm} | p{1.5cm} |}
\cline{2-7}
\multicolumn{1}{r|}{} & \multicolumn{3}{c||}{Inpatient beds} & \multicolumn{3}{c|}{Bed assign time} \\
\cline{2-7}
& 400\newline (Min.) & 500 \newline (Base) & 900\newline (Max.) & 1.8\newline (Min.) & 2.9 \newline (Base) & 4\newline (Max.) \\ \hline
Occupancy & 0.9185 \newline(+0.20\%) & 0.9167 & 0.7891 \newline (-13.92\%) & 0.89 \newline(+0.79\%) & 0.833 & 0.865 \newline(+2.04\%) \\
time, max diff & & 56 & & & 24 & \\ \hline
ED census & 59.35 \newline (-1.02\%) & 59.96 & 60.07 \newline (+0.18\%) & 56.72 \newline (-4.72\%) & 59.53 & 62.03 \newline (+4.20\%)  \\
time, max diff & & 139.5 & & & 19.625 & \\ \hline
ED boarders & 5.967 \newline (-1.06\%) & 6.031 & 6.043 \newline (+0.20\%) & 5.403 \newline (-7.94\%) & 5.869 & 6.238 \newline (+6.29\%)  \\
time, max diff & & 142.5 & & & 24 &  \\ \hline
Admit elect patients & 364.7 \newline (-19.95\%) & 455.6 & 479.3 \newline (+5.20\%) & 440.3 \newline (+0.39\%) & 438.6 & 437 \newline (-0.36\%)  \\
time, max diff & & 100 & & & 24 & \\ \hline
\end{tabular}
\end{table*}
%End of first table

\begin{table*}
\caption{Baseline sensitivity test results for transfer time and number of admitted elective patients}
\label{tab:3}       % Give a unique label
\begin{tabular}{l | r| r | r || r| r | r |}
%\begin{tabular}{l | p{1.5cm}| p{1.5cm} | p{1.5cm} || p{1.5cm}| p{1.5cm} | p{1.5cm} |}
\cline{2-7}
\multicolumn{1}{l|}{} & \multicolumn{3}{c||}{Transfer time} & \multicolumn{3}{c|}{Admit elect patients} \\
\cline{2-7}
& 0.5\newline (Min.) & 1.56 \newline (Base) & 2.5\newline (Max.) & 100\newline (Min.) & 150 \newline (Base)  & 200\newline (Max.) \\ \hline
Occupancy & 0.905 \newline(-0.43\%) & 0.909 & 0.912 \newline (+0.37\%) & 0.689 \newline(-17.59\%) & 0.836 & 0.917 \newline (+9.75\%) \\
time, max diff &  & 32 & &  & 16 & \\ \hline
ED census & 56.65 \newline(-4.84\%) & 59.53 & 61.82 \newline (+3.85\%) & 59.14 \newline (-1.37\%) & 59.96 & 59.98 \newline (+0.03\%) \\
time, max diff & & 19.5 & & & 139.5 & \\ \hline
ED boarders & 2.784 \newline (-53.28\%) & 5.959 & 8.628 \newline (+44.79\%) & 5.945 \newline (-1.42\%) & 6.031 & 6.033 \newline (+0.03\%) \\
time, max diff & & 23.125 & & & 142.25 & \\ \hline
Admit elect patients & 458.6 \newline (+0.64\%) & 455.7 & 453.2 \newline (-0.55\%) & 340.8 \newline (-17.82\%) & 414.7 & 455 \newline (+9.72\%) \\
time, max diff & & 48 & & & 16 & \\ \hline
\end{tabular}
\end{table*}
%End of second table
\end{landscape}

%\hline\noalign{\smallskip}

\begin{figure}[ht!]
  \includegraphics[width=0.8\textwidth]{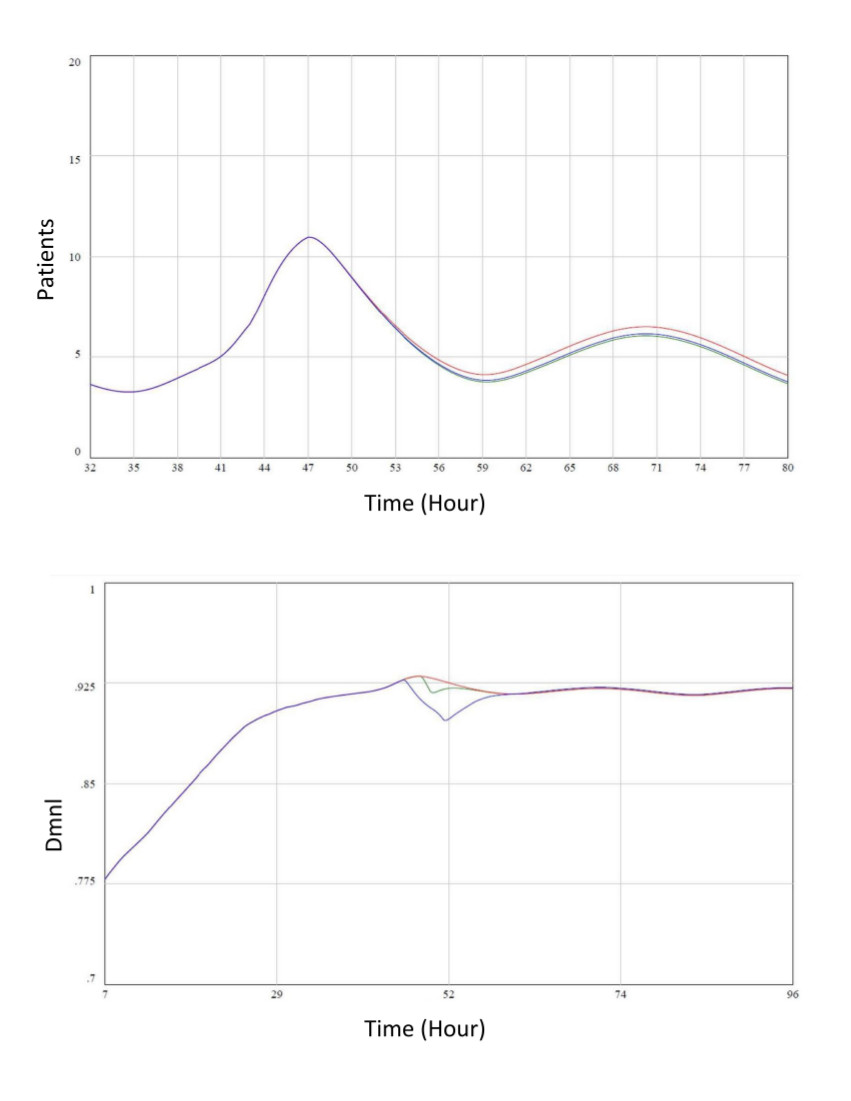}
\caption{The effect of changing the trigger threshold for ED boarders, measured on total number of ED boarders (upper graph) and occupancy rate (lower
graph).}
\label{fig:4}       % Give a unique label
\end{figure}

Two scenarios are used to compare the effects of changing parameter values. These are the baseline model, i.e. 'business as
usual' corresponding to what was experienced in October 2013 and the stressed model, where an exogenous surge of patients trigger
the protocol temporarily. The model shows some sensitivity towards a change in the trigger value for ED boarders. For instance,
changing the threshold from a default value of 10 to 7 yields a 5.69 \% increase in the number of ED boarders, due to the number
of returning patients while occupancy is lowered by 2.54 \% (see Figure \ref{fig:4}). Varying the ED census trigger had no effect
on the system. 

\begin{figure}[ht!]
  \includegraphics[width=0.8\textwidth]{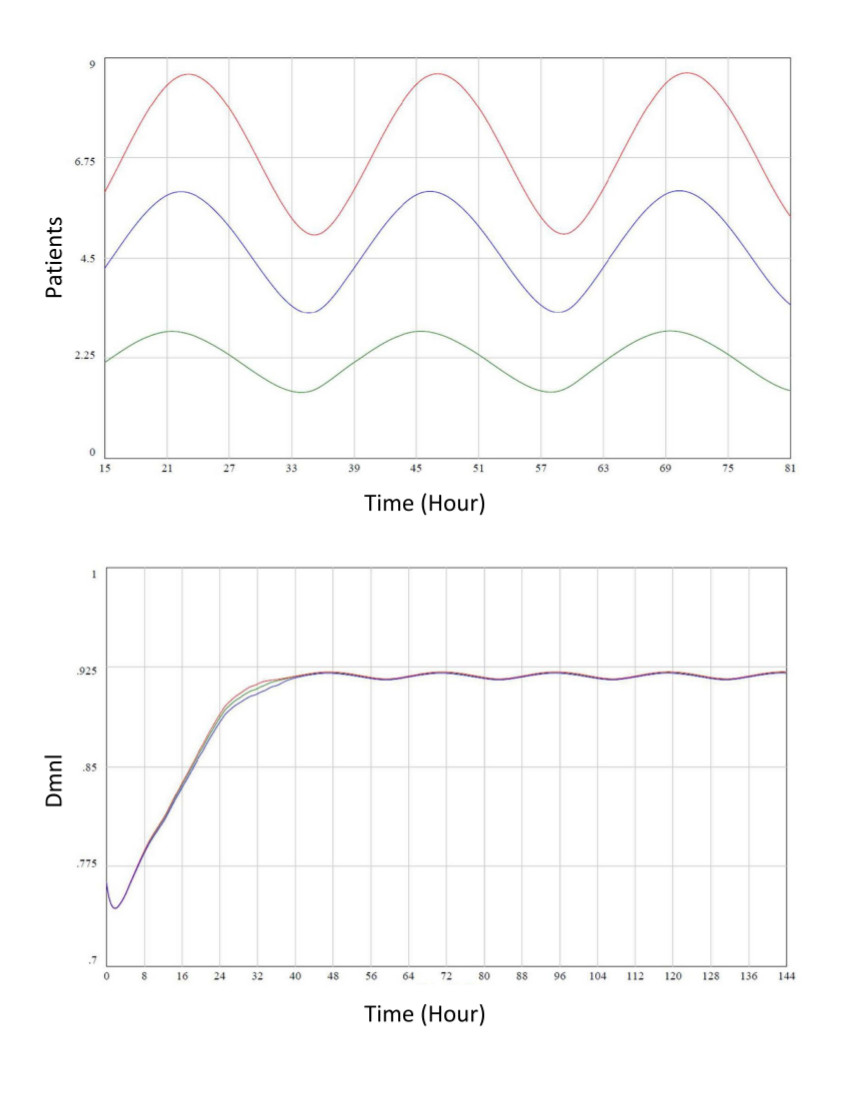}
\caption{The effect of changing transfer times on ED boarders (upper graph) and occupancy rate (lower graph). Lowering the transfer time both decrease
the number of ED boarders and the occupancy rate. Similarly, increasing the transfer time yields approximately the same effect in opposite direction. }
\label{fig:5}       % Give a unique label
\end{figure}

Switching the attention to the baseline model, four parameters were subject to analysis. These were the total inpatient capacity, bed assignment time,
transfer time, and mean incoming elective patients. In Figure \ref{fig:5}, the largest effect can be seen when changing the transfer time (53.28 \% and
44.79 \% respectively). 

Adjusting the inpatient capacity showed very little impact on the ED census (between 0.18 \% and 1.02 \%). The results obtained on both transfer time and inpatient
capacity are consistent with recently published studies applying other simulation techniques \cite{Khare2009,Hurwitz2014}. Being able to assign an inpatient bed faster shows some leverage
potential, implying enhanced communication between the receiving inpatient ward and the ED. Lastly, changing the number of incoming elective patients had little
impact on the ED census- and boarders, while occupancy rates and the number of admitted patients were impacted. 

\section{Discussion}
\label{discus}
The problem of ED crowding, which has been increasing despite the various policy recommendations for its reduction, can be described as a policy resistance problem, 
where system structure defeats the policy changes designed to improve its performance. System dynamics, through the quantification of system behaviour as derived from its structure, 
has been proven to be a suitable tool for the description and analysis of policy resistance problems. This study presents a system dynamics model capable of analysing the effects 
of a local protocol to alleviate ED crowding alongside other more generic parameters that have been identified in the literature as relevant levers for crowding. 
Three main results can be extracted from the study.         % Policy resistance and System Dynamics Model justification (DS)

First, the developed system dynamics model was validated both qualitatively, through iterative consensus-seeking consultations with emergency physicians on the process
being modelled, and quantitatively, through satisfactory replication of census data and exhibiting plausible behaviour when exposed to extreme conditions.

Next, the developed model was stressed so that the protocol was activated in order to analyse the effects of changing the trigger values for both ED census- and
boarders. The stressing of the model corresponds to a brief but significant elevation in ED patient arrivals, similar to the event of a major accident. What EDs normally experience is sustained levels of increased workload that eventually leads to fatigue and subsequent crowding. In the simulation, the effects are identical with a single difference being the requirement of a longer time window in the latter case. Having an ED census beyond the threshold limit is no rare incidence at the case hospital. The ED is routinely busy during the afternoon where patient arrivals have been high for several hours. Patients typically arrive faster than those either discharged directly or leaving for an inpatient ward. However, the
congestion is time limited since the patient arrivals will decrease, giving the ED a chance to recover. Furthermore, discharging patients around the clock is a followed practice at the case hospital \cite{Wong2010}. Therefore, the ED boarder trigger is the parameter that can
slow down the patient flow. If this threshold is lowered to 7 from 10, ED census increases 5.47 \% and ED boarders increase 5.69 \% while the occupancy rate and
number of admitted patients decrease by approximately 2.5 \%. A higher ED census is explained by the increased
amount of returning patients due to premature discharge from an inpatient ward. The effect is delayed since prematurely discharged patients will not present themselves to the ED immediately upon discharge.  

Finally, simulation of the baseline model yielded a number of other relevant findings. Fast transportation of ED boarders is identified as the main lever for
clearing ED congestion \cite{Shelton2010}. 

The current trends towards decreasing available inpatient beds should be evaluated in light of these effects. Bed assignment time was also revealed to be a
potential focus area for improving patient flow. When decreased to an average of 1 hour and 48 minutes (from a base of 2 hours and 55 minutes), ED census is
decreased by 4.72 \% while the occupancy rate is elevated by 0.79 \%. Inpatient bed capacity did not provide great leverage while changing the number of elective
patients primarily had an influence on admitted patients because of idle capacity.

The SD model can be extended in several ways. A benefit of system dynamics is its capability of modelling human behaviour. Social aspects of care and human
factors can therefore be included, for instance the performance effects of coping with increased workloads \cite{Morrison2011}. Some of the constants
included in the model can be investigated more in depth in future studies to reveal a more precise depiction of reality. One such example would be the relation between faster discharging times on
the probability of increased ED patient returns. Additionally, the ED patients that do return for more treatment may be at a higher risk of getting admitted once
more. Another possibility would be to focus on the transition between departments by explicitly modelling the receiving inpatient ward. Adopting a broader system focus would be an important first step in finding
sustainable efficiency improvements for the hospital as compared to looking at the ED in isolation \cite{Brailsford2004}.

\section{Limitations}
\label{limit}
The simulation model developed is only an approximation of reality as perceived by a limited number of ED physicians' mental models and is based on a limited data
sample. Reality is much more complex than the model presented, explaining the difference in displayed behaviour shown in Figure \ref{fig:3}. Patients are treated
as one large quantity despite being categorised according to their respective levels of acuity. Such differentiation could in the model be obtained by application of subscripts. 

The SD model introduces a varying inflow of elective patients with a fixed mean and standard deviation around the clock which represents an idealised scenario desirable from an ED perspective. The reality
is that most hospitals schedule elective admissions in batches typically in the mornings on Monday to Thursday. Elective patients getting admitted after surgery
are in need of inpatient beds at the same time that the daily wave of patients admitted from the ED are looking for beds. Furthermore, the hospital may have
policies as to admit either ED patients or elective patients in favour of the other in case of restricted inpatient capacity; something the model does not account
for.
 
The results generated from the developed SD model are dependent on the parameter representing the likelihood of a prematurely discharged inpatient returning to
the ED. This value was estimated based on current literature but may in reality be a variable that depend on how early the patient is discharged.
Since data stems from a single hospital, the model resembles one particular system. Generalizability potential is thus reduced. However, there are conclusions
that may be of interest to other hospitals with similar admission processes. The model's validity would be improved if similar data sets were acquired at other
emergency care hospitals. Finally, any interventional pilot studies that may follow the use of the proposed model would have to be assessed and compared to that
of the model's output. 

\section{Conclusion}
\label{conclude}
The main objective of this study was to investigate the effects of altering the trigger thresholds for a local protocol used for alleviating crowding situations
at a large tertiary care center and teaching hospital in Boston, MA. A side objective was to change the focus to more general parameters hypothesised as
alternative levers for ED crowding control. Through the development of a system dynamics model, it was shown that, for the protocol, changing the ED census
trigger had no effect on patient flow while the ED boarder trigger could speed up the patient flow when lowered at the expense of a slight increase in ED census.
More generally, faster bed assignment- and transfer times had a positive effect on the patient flow and are thus a recommended target area for potential process
improvements. This study underlines the capabilities of simulation as a feasible option in the search of sophisticated management decision support tools for
complex dynamic systems. 

\bibliographystyle{abbrv}
\bibliography{sys_dyn_ref}

\end{document}